\documentclass[
twocolumn,
]{ceurart}

\usepackage{graphicx}
\usepackage[export]{adjustbox}

\usepackage{url}

\begin{document}

\copyrightyear{2021}
\copyrightclause{Copyright for this paper by its authors.
  Use permitted under Creative Commons License Attribution 4.0
  International (CC BY 4.0).}

\conference{Fourth Workshop On Knowledge-Driven Analytics And Systems Impacting Human Quality Of Life (KDAH-CIKM-2021), November 01--05, 2021, Gold Coast, Queensland, Australia}

\title{PROVENANCE: An Intermediary-Free Solution for Digital Content Verification}

\author[1,2]{Bilal Yousuf}[%
orcid=0000-0001-6024-9084,
email=bilal.yousuf@adaptcentre.ie,
]

\author[1,2]{M. Atif Qureshi}[%
orcid=0000-0003-4413-4476,
email=muhammad.qureshi@adaptcentre.ie ,
]

\author[1]{Brendan Spillane}[%
orcid=0000-0001-5893-1340,
email=brendan.spillane@adaptcentre.ie,
]

\author[1]{Gary Munnelly}[%
orcid=0000-0002-7757-6142,
email=gary.munnelly@adaptcentre.ie,
]

\author[1]{Oisin Carroll}[%
orcid=0000-0001-9398-9388,
email=oisin.carroll@adaptcentre.ie,
]

\author[1]{Matthew Runswick}[%
orcid=0000-0002-0848-931X,
email=matthew.runswick@adaptcentre.ie ,
]

\author[3]{Kirsty Park}[%
email=kirsty.park@dcu.ie ,
]

\author[3]{Eileen Culloty}[%
orcid=0000-0001-7960-8462,
email=eileen.culloty@dcu.ie ,
]

\author[1]{Owen Conlan}[%
orcid=0000-0002-9054-9747,
email=owen.conlan@scss.tcd.ie ,
]

\author[3]{Jane Suiter}[%
orcid=0000-0002-2747-8069,
email=jane.suiter@dcu.ie,
]
    
\address[1]{ADAPT Centre, Trinity College Dublin}
\address[2]{ADAPT Centre, Technological University Dublin}
\address[3]{Institute for Future Media, Democracy and Society, Dublin City University}

\begin{abstract}
The threat posed by misinformation and disinformation is one of the defining challenges of the 21\textsuperscript{st} century. Provenance is designed to help combat this threat by warning users when the content they are looking at may be misinformation or disinformation. It is also designed to improve media literacy among its users and ultimately reduce susceptibility to the threat among vulnerable groups within society. The Provenance browser plugin checks the content that users see on the Internet and social media and provides warnings in their browser or social media feed. Unlike similar plugins, which require human experts to provide evaluations and can only provide simple binary warnings, Provenance's state of the art technology does not require human input and it analyses seven aspects of the content users see and provides warnings where necessary.
\end{abstract}


\maketitle

\section{Introduction}
Provenance is an intermediary-free solution for digital content verification to combat misinformation and disinformation on the Internet and social media. As per \cite{Rehm_2018}, it is designed to aid users by providing them with warning notifications in their browser or social media feed when viewing content that may be dangerous or problematic. The detailed warning notifications inform users which of the seven criteria Provenance's state of the art technology has detected an issue with and why. It significantly improves upon all known similar solutions in two ways. Firstly, existing solutions do not analyse the content the user is viewing and are thus limited to providing users with warnings based on the news agencies historical publication record and behaviour. Secondly, existing browser plugins only provide a single broad-spectrum warning about the content users are viewing whereas Provenance is capable of evaluating content under seven criteria and providing individual warnings for each. Provenance's warning notifications are also educational and designed to inspire users to be more cautious and critical of the information they consume. Thus, it will improve media literacy among users and make them less susceptible to the influence of misinformation and disinformation by making them more critical and reflective of the content they consume. 

There are significant research challenges in the design and development of Provenance. The main challenges include the huge volume of news and other content published each day, the combination of multimedia formats in each article or story, the high churn-rate and short shelf-life of news, and the fact that news content is often republished from wire services or from other publishers. These are compounded by the fact that misinformation and disinformation are often designed to masquerade as real news. Many disinformation sources share characteristics with the Lernaean Hydra of Greek mythology and re-post problematic content through multiple easy to set up websites or social media groups and reappear under different guises when they are identified and shut down. 

There are also a range of individual challenges within components of the Provenance platform. These include deriving a system to assign accurate writing quality scores for each piece of textual content, detecting when new facts introduced in a news article are indicative of disinformation or an evolution in an unfolding story, detecting image and video manipulations, or developing a system that can differentiate between anger and fear in disinformation and anger and fear in opinion news articles. There is also some difficulty in differentiating between news articles from alternative and independent agencies and news articles from disinformation sources due to often lower quality writing, more emotive content, and the reuse of images and videos.

This paper provides an update on the ongoing progress of developing Provenance. The remainder of this paper is organised as follows. Section \ref{Motivation} \textit{Motivation and Background} delves into the impetus for this project and situates it within other recent EU disinformation projects. Section \ref{Related Work} \textit{Related Work} provides a detailed overview of similar browser plugins and describes how Provenance advances the state of the art. Section \ref{Architecture Overview} \textit{Architecture Overview} contains system architecture diagrams and descriptions of each component in the Provenance platform. Section \ref{Provenance in Action} \textit{Provenance in Action} provides a detailed explanation of how the Provenance browser plugin provides warnings to the user. Section \ref{Use Cases} \textit{Use Cases} presents two use cases for the Provenance plugin to show in what scenarios we envision it being used.
Section \ref{Evaluation} \textit{Evaluation} briefly describes plans to evaluate the tool. Finally, section \ref{Conclusions} \textit{Conclusions} completes the paper with closing remarks.

\section{Motivation and Background} \label{Motivation} 
The proliferation of misinformation and disinformation on social media has been described as a strategic threat to democracy and society in the European Union (EU) \cite{European_Commission_2018, European_Commission_2017}. A recent EU study on the issue found that the common narratives of society \textit{"are being splintered by filter bubbles, and further ruined by micro-targeting."} \cite{Bayer_Bitiukova_2019}. The report points out that like a virus, misinformation and disinformation spread throughout society through social media and other platforms in open and closed groups to the detriment of democratic systems. This occurs when \textit{"Susceptible users become weaponized as instruments for disseminating disinformation and propaganda"} \cite{Bayer_Bitiukova_2019}.

The Presidents of the European Council, Commission and Parliament have all made increasingly public calls for concerted efforts to do more to combat the scourge of fake news to protect democracy. The President of the European Parliament has been the most forthright in this with a recent announcement that: \textit{"We must nurture our democracy \& defend our institutions against the corrosive power of hate speech, disinformation, fake news \& incitement to violence."} \cite{Ursula_von_der_Leyen_Twitter_fake_news}. As a result, the EU have funded a range of FP7, H2020 and other projects to combat misinformation and disinformation including WeVerify \cite{Aker_Sliwa_Dalvi_Bontcheva_2019, Marinova_et_al_2020}, SocialTruth \cite{Choras_2019}, PHEME \cite{Derczynski_Bontcheva_2014, Srijith_Hepple_Bontcheva_Preotiuc_Pietro_2017}, EUNOMIA \cite{Toumanidis_Heartfield_Kasnesis_Loukas_Patrikakis_2020} Fandango \cite{Martin_2020, Martin_2021} and the European Digital Media Observatory (EDMO) \cite{Ginsborg_Gori_2021}. Many other international organisations have also identified misinformation and disinformation as a threat and have increased efforts to combat it. These include the United Nations through its Verified platform \cite{United_Nations_Share_Verified} and the World Health Organisation \cite{World_Health_Organisation}. More can be read about these initiatives in the Poynter Institute's guide to national and international efforts to combat misinformation and disinformation around the world \cite{The_Poynter_Institute}. 

Provenance is a H2020 project\footnote{https://cordis.europa.eu/project/id/825227}, however it differs from many of the above as it is a user orientated intermediary-free solution to help consumers identify misinformation and disinformation as they browse the Internet and social media. It is also designed to improve media literacy skills by equipping consumers with the tools, knowledge and know-how to face this challenge now and into the future.

\section{Related Work} \label{Related Work} 
This review of related work will focus on comparable browser plugins designed to provide users with warning notifications about disinformation or other problematic content and which are currently active or maintained. The purpose of this review is to establish how Provenance advances the state of the art. 

\textbf{NewsGuard} \cite{NewsGuard} provides `nutrition' labels for news websites based on nine journalistic criteria. What differentiates it from many of the other fake news and bias detection browser plugins is that it does not use automated algorithms to assess news websites but rather relies on a team of journalists to conduct reviews. It comes as standard with Microsoft Edge, but a subscription is needed for other Internet browsers. Its notification icons appear as a browser extension in the upper right corner and within third party search engines and social media platforms. Clicking on its browser icon opens a nutrition label pane where users can quickly see whether the news website passes or fails any of the nine criteria. A link is also available for users to see a more detailed report. Visually, NewsGuard employs simple but effective white \checkmark on a green shield and red \textbf{x} iconography to denote when a website has passed or failed. NewsGuard's transparent methodology has resulted in their datasets being used for research \cite{Norregaard_Horne_Adali_2019}. While expert led analysis has its merits, it also has issues with scalability, personal biases, and response times. Aker also maintains that much of the credibility and transparency scoring provided by NewsGuard could be automated \cite{Aker_Kevin_Bontcheva_2019}.

\textbf{D\'ecodex} \cite{Le_Decodex} created by Le Monde originally started as an online search facility for users to check URLs against a list of known websites which spread misinformation and disinformation. They have since released a Facebook bot for users to directly chat to and a browser plugin that provides red, orange or blue notifications to denote whether a website regularly disseminates false information, whose reliability is doubtful, or if they are a parody website. When installed, the D\'ecodex icon becomes active when the website being viewed is listed in their database. It also produces a colour-coded popup with one of three standard warnings. Users cannot access detailed information about warnings, nor does it appear to be integrated with well-known search engines, social media platforms or discussion boards. D\'ecodex's allow/deny list approach means that scalability is difficult and the warnings it provides are based on the historical publication record of the website, not the content currently being viewed. Transparency is also limited. While still available, its development appears to be in stasis. 

\textbf{Media Bias Fact Check (MBFC)}\footnote{https://mediabiasfactcheck.com/} \cite{Media_Bias_Fact_Check_2021} is an extensive media bias resource curated by a small team of journalists and lay researchers who have undertaken detailed assessments of over 4000 media outlets. A transparent assessment methodology means that their datasets have been used for several research projects \cite{Kevin_Hogden_et_al_2018, Aker_Kevin_Bontcheva_2019}. Their team of researchers undertake in-depth analyses of news organisations and assess them using a standardised methodology, with some subjective judgement, to calculate a left/right bias score using their published formula. They also calculate scores for factual reporting and credibility. These reports are published on their website and updated from time to time. Each news website in their database is categorised as: left bias, left-centre bias, least biased, right-centre bias, right bias, pro-science, conspiracy-pseudoscience, fake news, or satire. While their browser extension conveys limited details, further information about each news source is available on their website. It draws on this dataset to inform users when they click on the notification icon as to which of these nine categories the news website they are viewing belongs to, including a brief explanation of the category. It also provides a link to the detailed MBFC report. The browser extension also provides Facebook and Twitter support by displaying a visual left/right bias scale on news articles that appear in users feeds with links to the MBFC detailed report and Factual Search\footnote{https://factualsearch.news} so that the user can investigate the topic further. While a valuable resource with considerable detail, MBFC's expert evaluations are based on the historical publication record of the news website and not an evaluation of the content the user is looking at. It is also a labour intensive and time consuming process. 

\textbf{Stopaganda Plus}\footnote{https://browserextension.dev/blog/stopagandaplus-helps-understanding-media-biases/}\cite{Stopaganda_Plus} is a browser extension that adds accuracy and bias decals to Facebook, Twitter, Reddit, DuckDuckGo and Google. These visual indicators extend the functionality of MBFC (who determine the scores) to these common information portals so that users may more easily choose high-quality information resources. It should be noted that this extension is not designed to provide users with detailed warning notifications when viewing a news website and thus is not directly comparable to the other systems or Provenance. It is included here due to its use of MBFC, the fact that it conveys limited visual information/warnings before the user visits an information source, and for plenitude. 

\subsection{No Longer Active}
Many other projects and services related to this work, which have been reviewed in the literature, c.f. \cite{Nordberg_Kavrestad_Nohlberg_2020, Hartwig_Reuter_2019, Gielczyk_Wawrzyniak_Choras_2019, Toumanidis_Heartfield_Kasnesis_Loukas_Patrikakis_2020, Skolkay_Filin_2019, Shu_Sliva_Wang_Tang_Liu_2017, Hanselowski_PVS_Schiller_Caspelherr_Chaudhuri_Meyer_Gurevych_2018}, now no longer appear to be active or working. This is concerning as despite the fact that misinformation and disinformation have been recognised as a threat to democracy and social cohesion, and the fact that browser plugins are one of the few citizen-orientated direct interventions which can help solve the problem at source while increasing long term media literacy, very few of the proposed solutions have been actively promoted or maintained. The main reason for this appears to be the fact that many of these plugins were developed by individuals or small teams, or even as part of a hackathon, and were thus lacked the resources to be actively maintained or updated to deal with changing technology such as browser updates or the rapidly evolving threats posed by misinformation and disinformation. The following present those related projects found in the literature, but which now no longer appear to be actively maintained, though some are still available to install. URLs have been included for posterity where possible as many do not have peer-reviewed publications.

\textbf{B.S Detector}\footnote{https://www.producthunt.com/posts/b-s-detector} relied on matching the URLs of content in the news feed to a known allow/deny list of sources of fake news and misinformation. 

\textbf{AreYouFakeNews.com}\footnote{https://github.com/N2ITN/are-you-fake-news} utilised Natural Language Processing (NLP) and deep learning to identify patterns of bias on websites. 

\textbf{Fake News Detector AI}\footnote{https://www.fakenewsai.com/} claimed to use a neural network to detect similarity between submitted URLs and known fake news websites.

\textbf{Fake News Detector}\footnote{https://fakenewsdetector.org/} was designed to learn from webpages flagged by users to detect other similar fake news webpages. 

\textbf{Trusted News}\footnote{https://trusted-news.com/} is a browser plugin that was designed to assess the objectivity of news articles. Its functionality was limited to `long form' news articles and it does not work with social media content.

\textbf{Fake News Guard}\footnote{http://fakenewsguard.com/} claimed to combine linguistic and network analysis techniques to identify fake news, however this can no longer be verified. 

\textbf{FiB}\footnote{https://projectfib.azurewebsites.net/} A browser extension built in a hackathon which was reviewed several times in the literature as a comparable system \cite{Goel_2016}.

\textbf{TrustedNews}\footnote{https://trusted-news.com/} Trusted News used AI to help users evaluate news articles by scoring their objectivity \cite{Trusted_News_Chrome_Extension_2020}. However, it does not work on social media and has issues with analysing webpages that require scrolling. 

\textbf{Trusty Tweet} \cite{Hartwig_Reuter_2019} was designed to help users deal with fake news tweets and to increase media literacy. Their transparent approach is designed to prevent reactance and increase trust. Early user evaluations showed promise. 

\textbf{Check-It} \cite{Paschalides_Christodoulou_Andreou_Pallis_Dikaiakos_Kornilakis_Markatos_2019} was designed to analyse a range of signals to identify fake news. It was focused on user privacy with computation undertaken locally. Their approach used a combination of linguistic models, fact checking, and website and social media user allow/deny lists. 

\subsection{Out of Scope Approaches}
Some misinformation and disinformation detection tools which have been reviewed in other papers have not been included in this literature review. This is because they are not a browser plugin or they are a paid for b2b service (Fakebox \cite{Machinebox_Fakebox_2021}; AreYouFakeNews \cite{Estela_2021}), they are focused on an aligned but separate issue e.g., detection of bias or detection of reused and or manipulated images (Ground.News \cite{Ground_News}; SurfSafe \cite{Bhat_2021}), they are specifically for fact checking (BRENDA \cite{Botnevik_Sakariassen_Setty_2020}, CredEye \cite{Popat_2018}), they have pivoted into a B2B platform (FightHoax \cite{FightHoax}), they are not user orientated (Credible News \cite{Hardalov_Koychev_Nakov_2016, Hardalov_2019}), or they are research systems and have not been made available to the public \cite{Hanselowski_PVS_Schiller_Caspelherr_Chaudhuri_Meyer_Gurevych_2018, Zhou_Jain_Phoha_Zafarani_2020}. While relevant to combating disinformation, these are not directly comparable to Provenance.

\subsection{Advancing the State of the Art}
This review demonstrates that browser plugins are a common user-orientated approach to combat misinformation and disinformation. However, Provenance adopts a significantly more advanced and granular methodology than current or previous efforts in the domain. The warnings provided by earlier plugins are often based on the news website's history of publishing misinformation and disinformation. Thus, they are limited to providing a coarse-grained retrospective analysis of the news website's publication history. In contrast, Provenance's fine-grained approach is designed to analyse the content of the news webpage or users' social media feeds and, where necessary, provide an easy to understand warning to the user when the content they are viewing may be problematic or symptomatic of disinformation. In the cases where linguistic analysis or other machine learning approaches have been utilized, the results are not presented to the user in an explainable or transparent way. Some of these methods have also proven susceptible to adversarial attacks, whereby text may be augmented slightly to fool pretrained models \cite{Zhang_Sheng_Alhazmi_Li_2019, Zhou_Guan_Bhat_Hsu_2019}.

Two factors differentiating Provenance from the plugins described above are their limited reach and scalability. Many of the above plugins do not provide any information for some heavily trafficked news websites such as the LA Times, Al Jazeera, and the Independent.co.uk. This is likely due to limiting factors of time and labour of including humans in the disinformation judgement process. While no one doubts the benefits of highly trained expert judgement, the size and nature of the rapidly evolving media landscape, especially in regard to misinformation and disinformation in which publishers are prone to rapid growth, failure and re-branding, means that providing human ratings is a never ending game of whack-a-mole. Current solutions are only partially succeeding in providing judgements of some news agencies. None have attempted to analyse the millions of pieces of content they publish daily. Unlike each of the plugins described above, Provenance does not require a human-in-the-loop, nor does it need to be backed by human-generated allow/deny lists. Its architecture supports fully automated and intermediary free analysis of news content. 

The ability to evaluate news articles against seven criteria and provide users with visual notifications and deeper explanations is also a significant advancement on the state of the art and a direct benefit to users in three ways. First, and most importantly, users will be made aware of individual issues with the content they are consuming and can thus decide whether they will continue viewing it or look for alternative sources. Second, it will help develop users' media literacy skills by making them aware of the different caution worthy indicators and how to check them, making them less susceptible to misinformation and disinformation in the future. Third, the nature of these systems means that they cannot be properly examined. In contrast, a full description of Provenance's system architecture is provided below. It is also currently undergoing evaluation and testing and the results will be published in time.







\begin{figure*}
\framebox{
\includegraphics[width=14.3cm,height=\textheight,keepaspectratio]{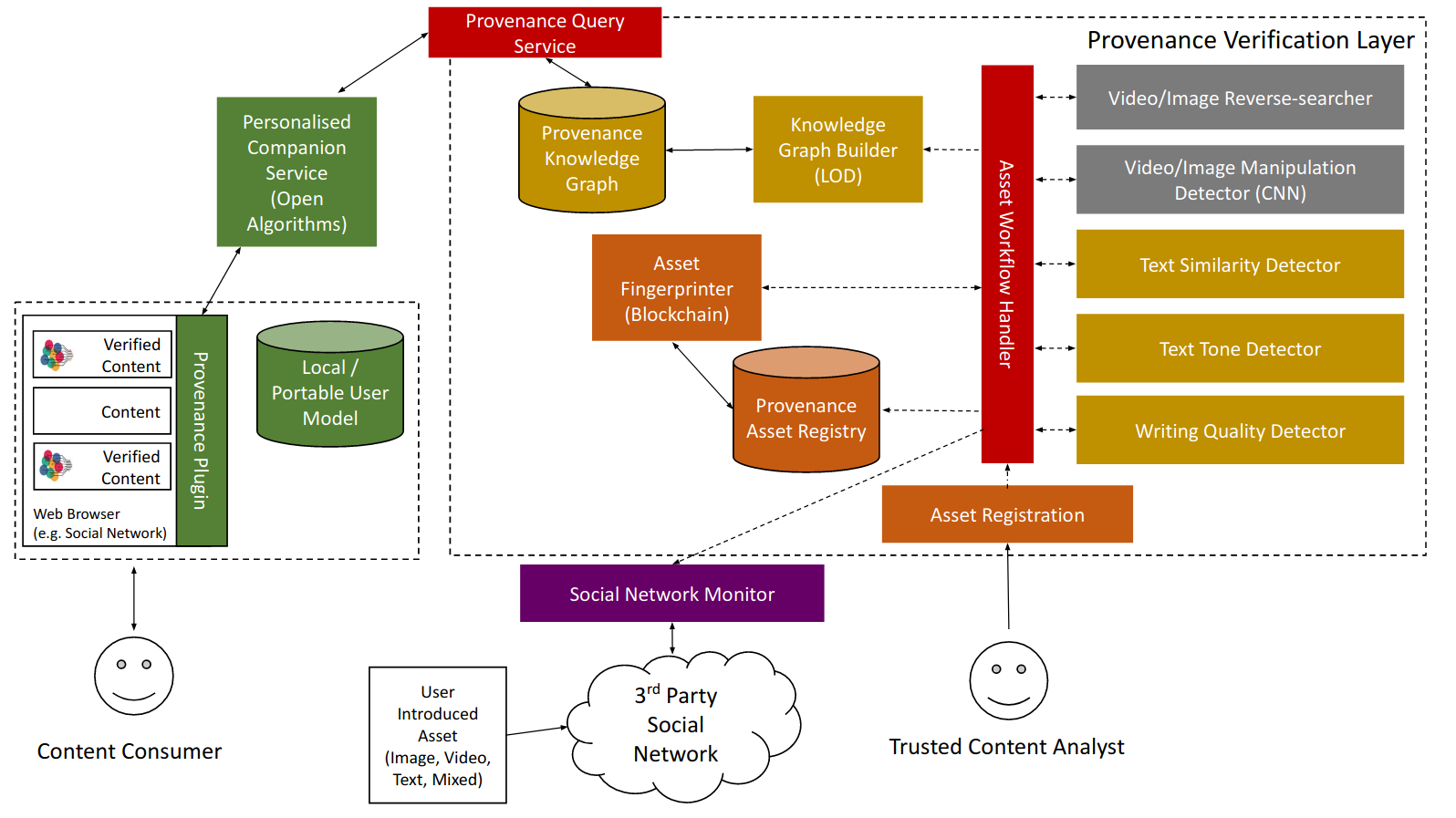}}
\caption{Provenance System Architecture: Dashed lines denote REST API calls, solid lines denote local access.}
\vspace{-1.3em}
\label{fig:architecture}
\end{figure*}

\section{Architecture Overview} \label{Architecture Overview}
The system architecture for Provenance is shown in Figure \ref{fig:architecture}. The components and services use REST APIs serving JSON for easy, reliable, and fast data exchanges across internal subsystems.

Data in the form of webpages or social media content is ingested by Provenance either through the \textit{Social Network Monitor} or by a \textit{Trusted Content Analyst} (e.g., a journalist or fact checker). The \textit{Social Network Monitor} service discovers content using NewsWhip's\footnote{https://www.newswhip.com} social network monitoring platform. The introduced asset is enriched with social engagement data (e.g., likes and shares) and is forwarded to the \textit{Asset Workflow Handler} service.

The \textit{Asset Workflow Handler} separates the incoming data (e.g., a news webpage) into individual assets such as images, video, text, etc. These assets are registered with the \textit{Asset Fingerprinter} before being disseminated to the analytical components (\textit{Video/Image Reverse-searcher}, \textit{Video/Image Manipulation Detector}, \textit{Text Similarity  Detector}, \textit{Text Tone Detector}, and \textit{Writing Quality Detector}) to determine if they exhibit any features which normally characterise misleading, questionable, or unsubstantiated information. The output of each analytical service, and the initial data passed from the \textit{Social Network Monitor} are combined and sent to the \textit{Knowledge Graph} where they are stored.

The Knowledge Graph may be queried by the \textit{Provenance Query Service} to retrieve the results of analysis for a given webpage. The Provenance plugin, installed in the user's browser, leverages this query service to retrieve information about webpages that a user is currently viewing. If the webpage has been analysed by Provenance, and exhibits questionable features, the plugin will issue a warning to the user, indicating that they may want to further investigate the claims made in the article's content. The \textit{Personalised Companion Service} is used to determine how this information should be presented for an individual user.

\subsection{Key Components}
\subsubsection{Social Network Monitor}
The \textit{Social Network Monitor} communicates with NewsWhip’s \textit{Social Network API} to identify assets which should be ingested by Provenance. Finding assets involves querying Newswhip's API with a parameterized search request. The call to NewsWhip’s \textit{Social Network API} is automatically invoked periodically to maintain an updated record of trending news articles and social media posts. Assets detected by NewsWhip are enriched through social scoring. The URL, titles, summaries, images and videos (if any), along with the enrichment data, is extracted from the article and provided to Provenance. Assets composed only of text, for example, are registered in fragments consisting of news feed/article title, the summary, and user engagement data.

\subsubsection{Asset Registration}
A dedicated \textit{Asset Registration} web interface also allows \textit{Trusted Content Analysts} to add assets into the \textit{Asset Workflow Handler}. \textit{Trusted Content Analysts} are stakeholders such as journalists and other representatives of news agencies and wire services, fact checkers, debunkers, and original content creators who may want to register their multimedia content assets. In future, this facility will be made more widely available to allow the general public to send content directly to Provenance. It may also be integrated with news publication platforms and content management systems so that content is automatically added. The primary task of this component is to enable third-parties to register assets that have not been discovered by the \textit{Social Network Monitor}.

\subsubsection{Asset Workflow Handler}
The \textit{Asset Workflow Handler} is the component of the \textit{Provenance Verification Layer} that is responsible for orchestrating the components and data within the layer. This component's primary task is to distribute assets to different components for further processing. It invokes the service interfaces and handles the data flow between the services. By utilising the \textit{Asset Workflow Handler}, components are loosely coupled, thus mitigating direct component-to-component communications. This will enable Provenance to work with the variety of APIs exposed from the existing tools/components. Moreover, the APIs can be adjusted to meet Provenance’s specific needs. Due to this modular design, new components can be easily added to the \textit{Provenance Verification Layer} (e.g., detection of bias \cite{Spillane_Lawless_Wade_2020}, tabloidization \cite{Spillane_Hoe_Brady_Wade_Lawless_2020}, and hate speech \cite{Schmidt_Wiegand_2017}), and connected to the \textit{Asset Workflow Handler}.

\subsubsection{Video/Image Reverse Searcher}
The \textit{Video/Image Reverse Searcher} is a key component for creating a large-scale annotated dataset for detecting manipulated visual content. The dataset consists of three distinct parts. The first part includes 45,000 images, each captured by a unique device (i.e., 45,000 different cameras have been used). Half of these images are real, and the other half has been digitally manipulated by applying a random image processing operation to a local area of the image. Since the sensor pattern noise present in images is unique to each sensor (i.e., camera), this dataset introduces large diversity, such as noise. The second part of the dataset uses imaging software in cameras to introduce a large diversity of artefacts in images. Commonly available camera brands and models were identified and used to collect a dataset of 50,000 images. Half of these images were digitally manipulated using an advanced image editing method based on Generative Adversarial Networks (GAN) \cite{Goodfellow_2014}. Finally, the third part of the dataset consists of 2,000 images downloaded from the Internet representing “real-life” (uncontrolled) manipulated images created by random people. For all of the manipulated samples collected for the third part of the data, the matching unmanipulated image was also collected. This component's primary task is to enable search operation for videos and images.

\subsubsection{Video/Image Manipulation Detector}
The Provenance \textit{Video/Image Manipulation Detector} identifies if an image or video has been manipulated in comparison to its source. This work is based on the PIZZARO\footnote{http://zoi.utia.cas.cz/node/180/0459504} project. It utilises recent developments achieved by deep learning-based methods to enable an instant detection of manipulations in visual content. In addition, use of the latest technologies based on Convolutional Networks will lead to tangible enhancements in integrity verification in visual content. The \textit{Video/Image Manipulation Detector} increases trust and improves governance. The solution is designed to build a web-based system to assess visual content in a real-world setting. The \textit{Video/Image Manipulation Detector} will further support the development of user skills in detecting false visual information themselves by providing a world-class image forensic technology. The \textit{Video/Image Manipulation Detector} has a special focus on developing a solution that will be intuitive and easy to understand and interpret for end-users, thereby increasing its uptake by the public and its impact on the information system. This component's primary task is to detect if the image and video are manipulated by comparing them with previously registered images and videos in the system.

\subsubsection{Asset Fingerprinter and Asset Registry}
The \textit{Asset Fingerprinter and Asset Registry} provide traceability of registered content. It is based on Blockchain technology, making content immutable and enabling the verification of the sources and alterations to the content. Registered assets are handed to the \textit{Asset Fingerprinter} via the \textit{Asset Workflow Handler}. Due to the General Data Protection Regulation (GDPR) and the size of some assets, the hash of the data is stored on Blockchain. Azure Storage is used as the Blockchain, and the assets themselves, including large files, are stored using an off-line storage service available to store multimedia files. Blockchain is used due to its innate data integrity which is important to prove the traceability of registered content if the tool was ever targeted as part of a combined disinformation and hacking campaign. This component's primary task is the traceability of registered content via Blockchain.

\subsubsection{Text Similarity Detector}
News is regularly republished nationally and locally  from international wire services such as Reuters, Agence France-Presse (AFP) and Associated Press (AP). In a bid to lower costs, many news agencies who are not in competition negotiate deals to republish each other's content. Similarly, less trustworthy news outlets often put `spins' on existing articles, where correct articles are modified to contain false information.

To combat this, the \textit{Text Similarity Detector} in Provenance attempts to verify the textual content of an article by comparing it to similar articles published elsewhere. A backlog of trustworthy articles is stored in an Elasticsearch database with a BM25 similarity index \cite{robertson1994some}. As BM25 under-performs with very long documents \cite{Lv_Zhai_2011}, only the title and first 10 sentences are used in the index. Once similar articles have been found the component searches for facts given in the query article in the similar ones. Facts in an article are found by taking sentences with a low subjectivity from TextBlob's sentiment analysis model \cite{Loria_2020}. The similarity of two facts is the cosine similarity of the vector embedding of both, which is provided by Google's multilingual text model \cite{yang2019multilingual}. If enough of the article's factual content cannot be verified, the plugin displays a warning.

\subsubsection{Text Tone Detector}
Intuitively, one would expect that impartial news sources would use impartial, unemotive language to convey the facts of a story. Recent research has shown that emotions such as fear, anger, sadness, doubt, and the absence of joy and happiness are indicative of misinformation and disinformation \cite{Parikh_Patil_Atrey_2019, Paschen_2019, Zhang_Cao_Li_Sheng_Zhong_Shu_2021}. Provenance's \textit{Text Tone Detector} is designed to identify emotions in text which may indicate that the news source is unreliable. Threshold values are used to determine whether caution should be shown, and the degree of caution is determined by how far the calculated value deviates from the threshold value.

\subsubsection{Writing Quality Detector}
Provenance's \textit{Writing Quality Detector} computes a writing quality score (WQS) for the textual content the user is viewing and provides a warning when it falls below a threshold value. Writing quality is closely related to cohesion and coherence \cite{Singh_2020}. Within the context of news, high quality writing is indicative of paid professional journalism from mainstream, independent, and to a lesser degree, alternative news agencies, whereas low quality writing is indicative of amateur or unprofessional news production processes \cite{Chung_Kim_2021}. This high/low quality differentiation is also apparent in other domains such as academia, publishing, commercial, and blogs and information websites. While NLP techniques exist to derive writing quality \cite{Klyuev_2018}, and others have called for it to be used to identify misinformation and disinformation \cite{Spradling_Straub_Strong_2021, Fuhr_2018}, only two examples of systems could be found in the literature which actually calculate writing quality \cite{Connie_Fan_2017, Jo_Muhamed_Nuthakki_Singhania_2018}.    

To calculate WQSs for Provenance, a dataset of news articles, blog posts, and other website content, much of which had characteristics symptomatic of disinformation, was annotated in a crowdsourced study to identify terms and phrases indicative of low quality writing. A WQS for each piece of content was then derived using a standard formula. This was subject to testing and expert evaluation to ensure the WQS the formula produced accurately reflected each piece of content. Models were then trained on the dataset which showed that the WQS could be automatically generated with a high degree of accuracy. These models and the overall process are currently undergoing formal evaluation.

\subsubsection{Knowledge Graph and Knowledge Graph Builder}
\label{sec:kg-and-kg-builder}
The Provenance \textit{Knowledge Graph} stores a record of all the articles introduced to Provenance via the \textit{Social Network Monitor} service or via Asset Registration from a \textit{Trusted Content Analyst}. It is also a record of all analysis performed on said assets.

\begin{figure}
  \centering
  \fbox{\includegraphics[width=6.85cm]{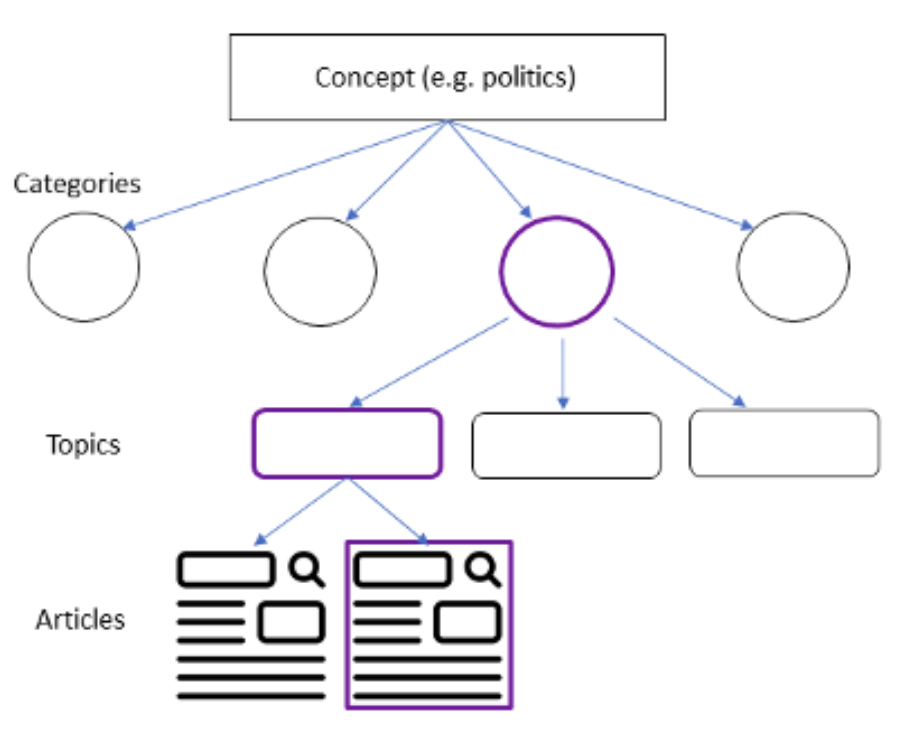}}
  \caption{Knowledge Graph categorisations of assets.}
  \vspace{-1.3em}
  \label{fig:Knowledge_Graph}
\end{figure}

The content is organised according to concept, categories and topics. For example, a news article discussing politics can be categorised according to the left/right political spectrum followed by the topics discussed as shown in Figure \ref{fig:Knowledge_Graph}. Each node at the article level is split according to text, image and video. 

The output of the \textit{Video/Image Reverse Searcher} includes the N most similar images/videos, distance measures and geometric validation results. The data from the \textit{Video/Image Manipulation Detector} includes the probability of manipulations and the area of polygons. These are sent as JSON objects to the \textit{Knowledge Graph} where they are stored as entities in a triplestore.

Modelling of Provenance data is achieved using a combination of the RDF Data Cube vocabulary \cite{world2014rdf} to store statistical information such as the outputs from the various analytical components, and the Dublin Core/BIBO vocabularies \cite{dublin2012dublin} to model bibliographic information about the assets themselves. Some use is also made of the FOAF\footnote{\url{http://xmlns.com/foaf/spec/}} vocabulary to model information such as content publishers, which are naturally represented as \texttt{foaf:Agent} entities.

The \textit{Knowledge Graph Builder} is responsible for exposing a REST API which the \textit{Asset Workflow Handler} may use to upload assets as JSON, and then transforming the JSON into triples which are stored in a triplestore. In Provenance, this is achieved using JOPA \cite{ledvinka2015jopa}: a Java library which can be used to map POJOs to triples. Using Spring Boot\footnote{\url{https://spring.io/projects/spring-boot}}, a REST API accepting JSON is exposed. The uploaded JSON is serialized into POJOs using Spring Boot's built-in version of Jackson. JOPA is then used to serialize the triples out to an RDF4J\footnote{\url{https://rdf4j.org/}} instance.

The same serialization process works in reverse, allowing the \textit{Provenance Query Service} to expose both a JSON REST endpoint which can produce JSON objects from the results of a canned SPARQL query exposed via a Spring Boot REST endpoint, and a much lower level raw SPARQL endpoint from the triplestore, for those who want a high level of control over their queries.

\subsubsection{Provenance Query Service}
The \textit{Provenance Query Service} is the interface to the \textit{Verification Layer} and offers external trusted services with the means to request verification information about a webpage or article. It will also allow trusted services with a means to identify the relatedness of content (through similarity and the \textit{Knowledge Graph}) and determine if content has been modified. As the results of all analysis are stored in the \textit{Knowledge Graph}, the \textit{Provenance Query Service} is effectively a proxy between the user-facing front-end, and the query interface to whatever storage medium is used to implement the Knowledge Graph.

As mentioned in Section \ref{sec:kg-and-kg-builder}, the Provenance Query Service exposes both a raw SPARQL endpoint and a REST API which provides endpoints for a number of canned SPARQL queries which return JSON objects. It is envisioned that the vast majority of user cases will be covered by the REST API, making it easier for developers to access data that is helpful to users. However, it is worthwhile to allow lower level access to the KG's contents in the event of unforeseen requirements being placed on the KG.

\subsubsection{Personalised Companion Service }
The \textit{Personalised Companion Service} manages the Provenance verification indicator, the minimal user model, and user scrutability and control. The verification indicator is implemented as a Chrome Extension and works on the Facebook and Twitter platforms and with articles published by news agencies. The \textit{Personalised Companion Service} uses the user’s interests, domain knowledge, digital literacy, and the warning preferences stored in the Minimal User Model to determine whether to highlight caution or show the verification indicator without caution. The \textit{Personalised Companion Service} uses the data provided by the \textit{Asset Fingerprinter}, the \textit{Video/Image Reverse Searcher} and \textit{Video/Image Manipulation Detector}, and the \textit{Text Similarity, Tone and Writing Quality Detector} components to create the set of icons that are presented to users, who can explore the levels of verification presented through the visual iconography.

\section{Provenance in Action} \label{Provenance in Action}
\begin{figure}[htbp!]
  \centering
  \fbox{\includegraphics[width=6.9cm]{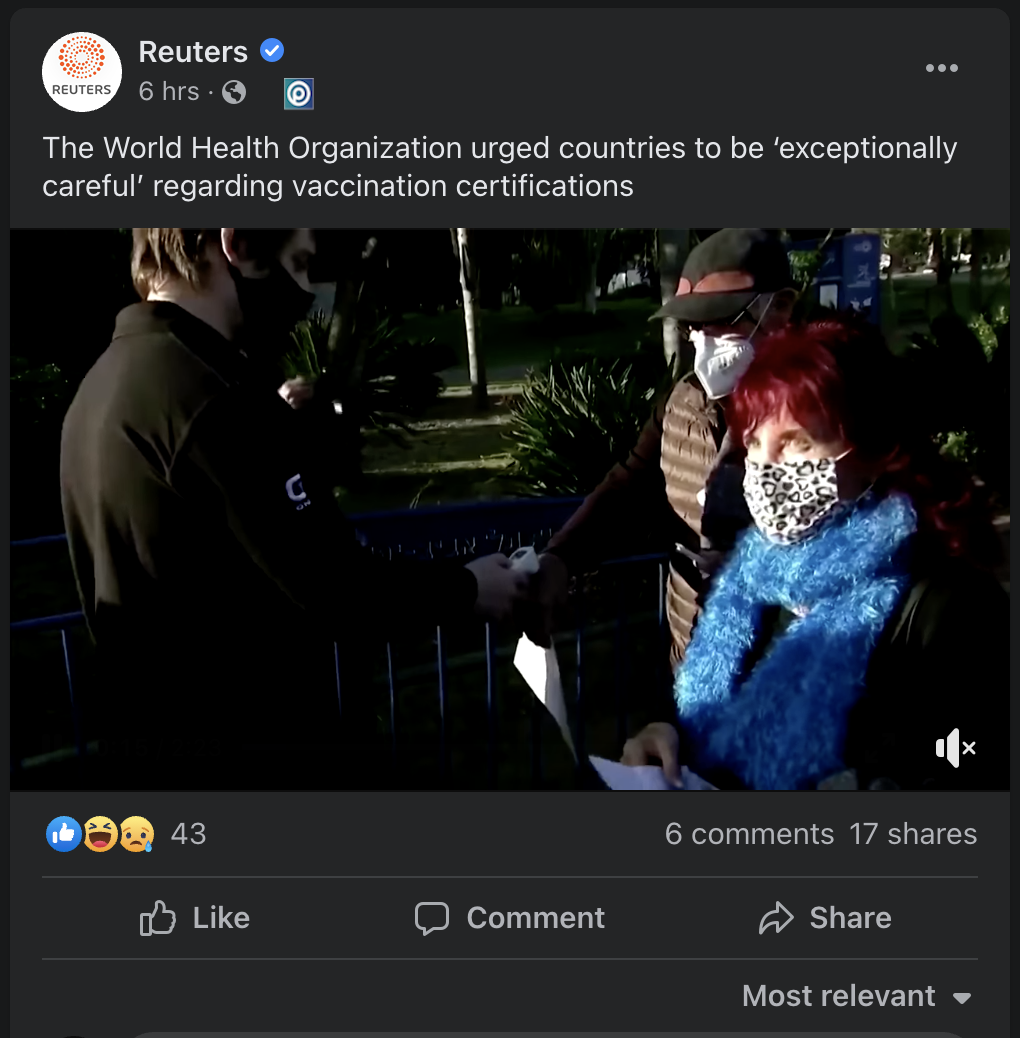}}

  \caption{A user's Facebook feed showing the Provenance icon in blue indicating that there are no warnings.}
\vspace{-1.2em}
  \label{fig:no_warning}
\end{figure}

\begin{figure}[htbp!]
  \centering
  \fbox{\includegraphics[width=6.9cm]{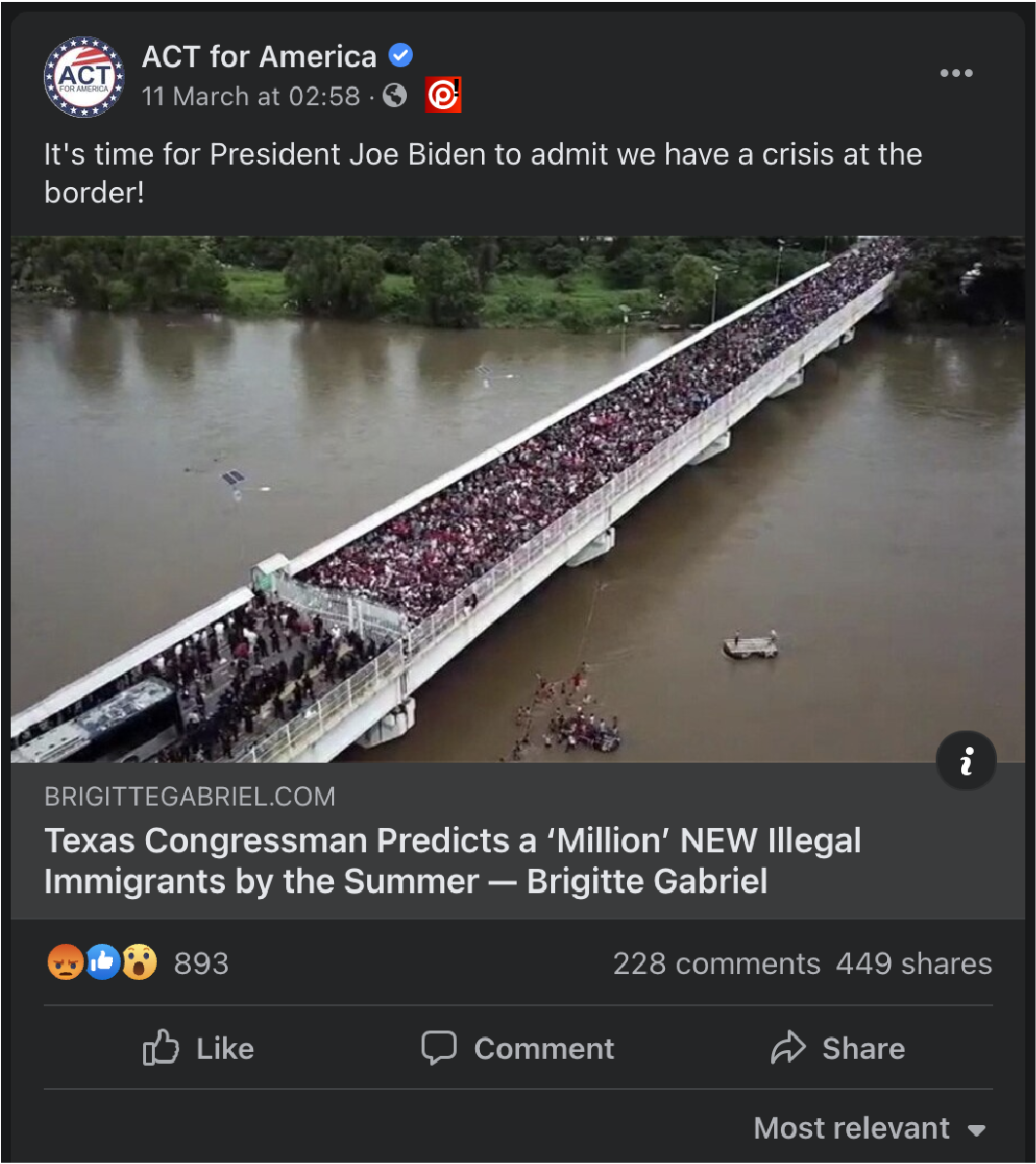}}

  \caption{The Provenance icon in red (with exclamation mark) indicating that this article has one or more issues which are worthy of caution.}
\vspace{-1.2em}
  \label{fig:first_warning}
\end{figure}

\begin{figure}[htbp!]
  \centering
  \fbox{\includegraphics[width=6.9cm]{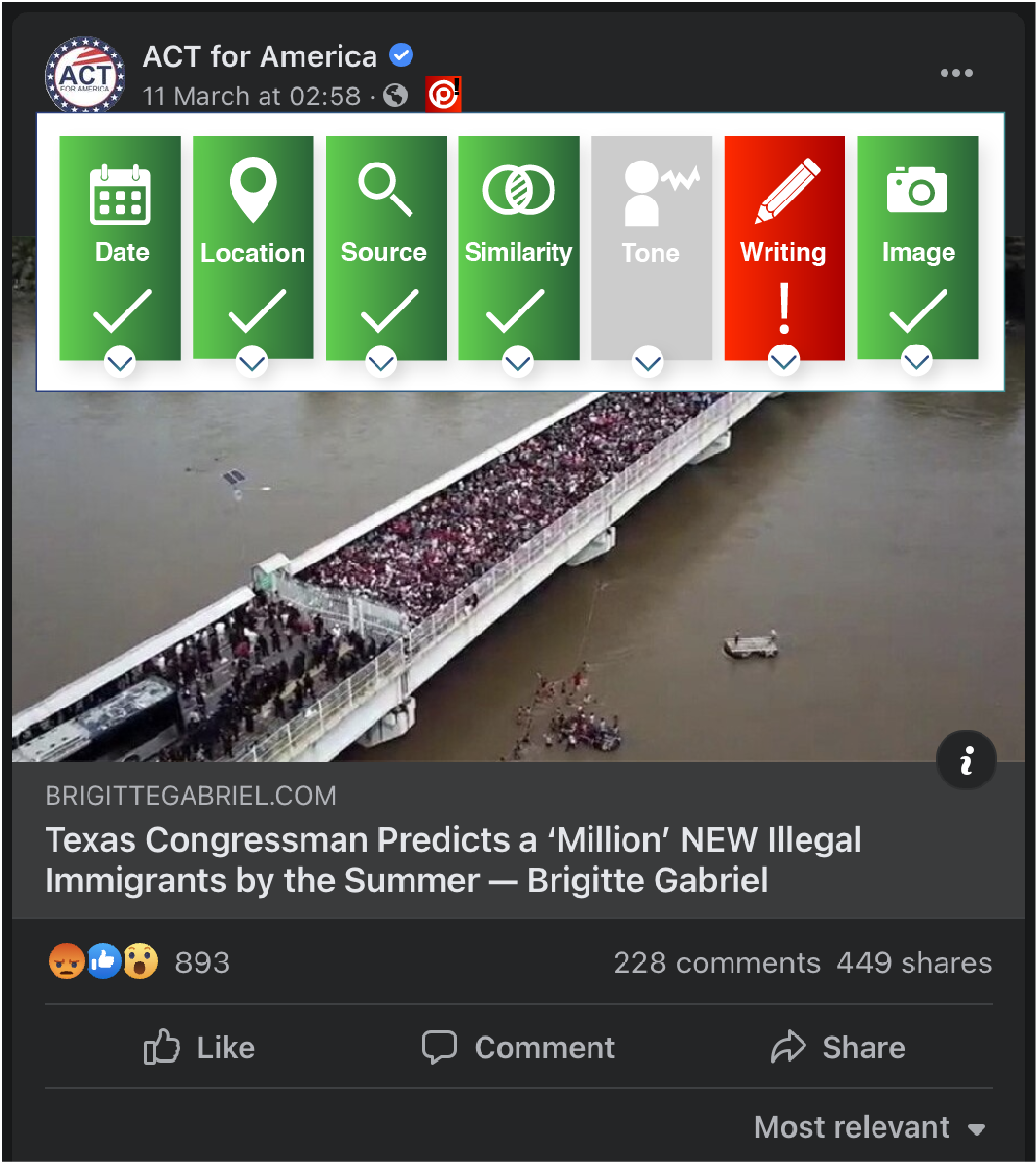}}

  \caption{An initial explanation pane appears when then user clicks on the Provenance icon in their social media feed.}
\vspace{-1.2em}
  \label{fig:expanded_warning}
\end{figure}

\begin{figure}[htbp!]
  \centering
  \fbox{\includegraphics[width=6.9cm]{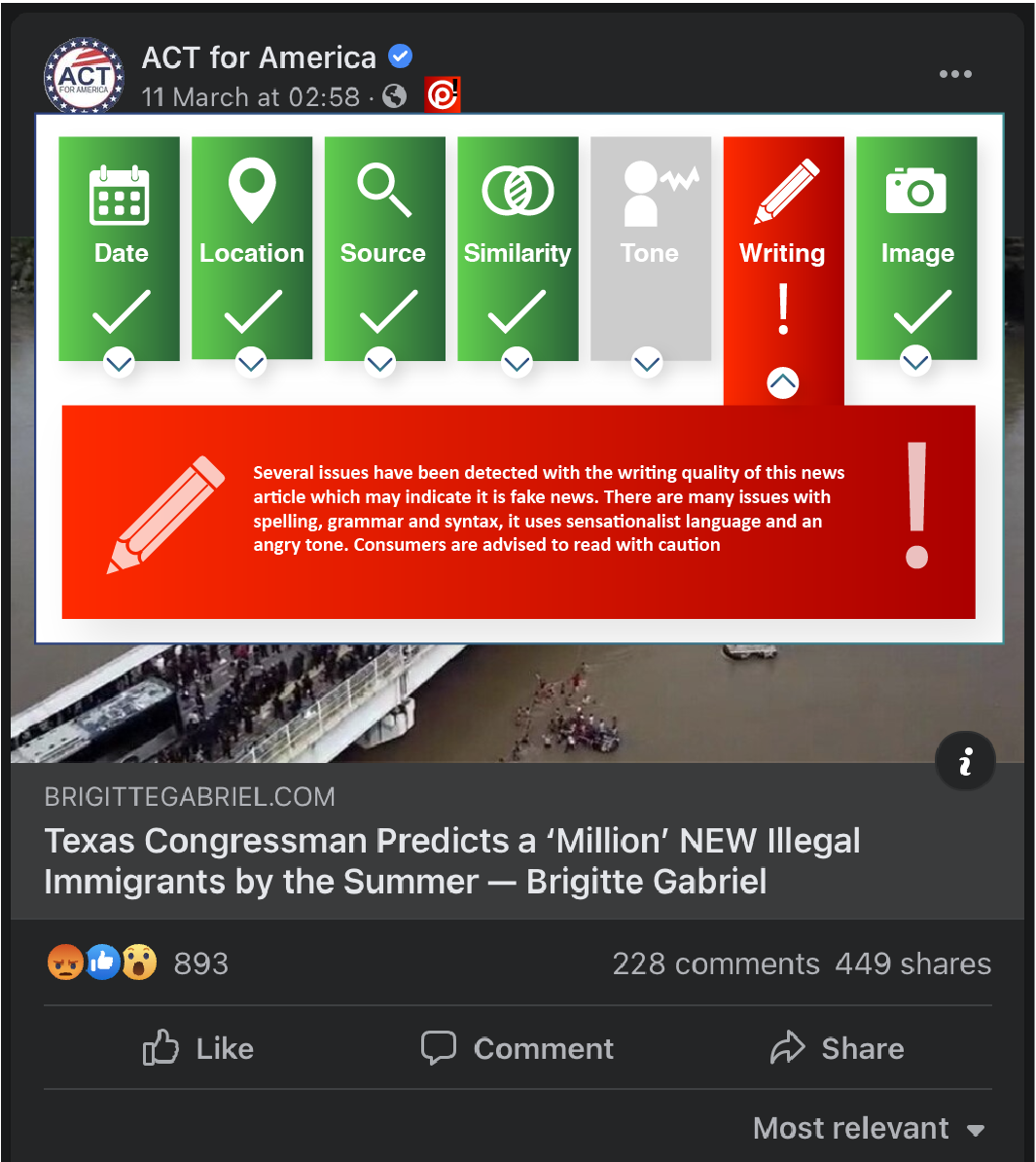}}

  \caption{A detailed explanation pane appears when the user clicks on any of the seven categories Provenance analyses the news item under.}
\vspace{-1.2em}
  \label{fig:detailed_warning}
\end{figure}

The Provenance browser plugin is designed to provide users with easy to understand, granular and cautionary warnings about the content they are consuming. These warnings are provided via an in-browser icon beside the address bar when the user is browsing the Internet, or within their Facebook and Twitter social media feeds beside the content they are viewing. Figures \ref{fig:no_warning} - \ref{fig:detailed_warning} show how Provenance and its visual warnings appear to a user - who has the Provenance plugin installed - within their Facebook social media feed. The Provenance icon appears as a small blue square with a white P above each content item that it has checked. When the icon background turns red (with a small exclamation mark), it indicates to the user that the content item is worthy of a cautionary warning. The following presents the four main states of Provenance which a user will see.

Figure \ref{fig:no_warning} shows a user's Facebook feed who has the Provenance browser plugin installed. The Provenance icon is visible at the top of each news article in the user's feed. In this image, the icon is blue which indicates that there are no warnings with this particular news item.

In Figure \ref{fig:first_warning}, the background of the Provenance icon within the user's news feed has turned red to indicate that this news item is worthy of one or more cautionary warnings. A small black exclamation mark has been added to the top right of the icon for colour blind users. 

In Figure \ref{fig:expanded_warning}, the user has clicked on the red Provenance icon. A window has appeared beneath the Provenance icon to show the user which of the seven criteria the news article was checked against that Provenance has detected an issue with. In this example, the red background and exclamation mark beneath the \textit{Writing Quality} icon indicates that this aspect of the news article is worthy of caution. The user may click on the downward arrow beneath each icon for further information. In this example, the \textit{Tone} icon is greyed out indicating that this could not be assessed by Provenance in this instance.

Figure \ref{fig:detailed_warning} shows a detailed explanation of the \textit{Writing Quality} warning after the user clicked on the option to expand it. It contains further information about how \textit{Writing Quality} score is calculated and why low quality writing is indicative of misinformation and disinformation.

\section{Use Cases: Provenance Plugin} \label{Use Cases}

\subsection{Social Media Timeline} 
On the recommendation of a friend, Mary installed the Provenance browser plugin due to increased concerns about the spread of misinformation and disinformation. The instructional video on the Provenance Chrome Extension webpage explained that Provenance uses seven criteria to verify digital content on the Internet and social media feeds. After installing the Provenance plugin, she notices that the news items in her Facebook timeline now display the Provenance icon beside the publisher's name. For most of the news stories, the Provenance icon shows a white P inside a white circle on a blue background. When she clicks on the blue Provenance icon, it opens a notification pane showing the seven verification criteria, all of which display a green background with a white \checkmark. 

She is able to click on each of the seven verification icons to read a detailed explanation for each criterion, why failing the criterion is an indication that the webpage or social media post may be misinformation or disinformation, and how the warning is derived. As all of the icons are green, she is reassured about the origin, veracity and overall quality of the news article. For some news items displayed on her timeline, she notices that the blue background of the Provenance icon has turned red. When she clicks on it, the same information pane displaying the same verification criteria appears, except one or more of the seven verification criteria now display a red background with an exclamation mark beneath. When she clicks on these, an additional detailed explanation pane appears underneath them to explain why it has failed. Reading through each warning including their detailed description, she gains a better understanding of how to identify misinformation and disinformation. In both instances, Mary has become more aware of the need to critically check the news she consumes and more aware of good media literacy habits in general.

\subsection{News Websites} Mary regularly visits news websites to inform herself of current affairs. Usually, the Provenance icon, which is visible to the right of her browser's address bar, displays a white P inside a white circle on a blue background. However, recently when she was visiting news websites to read more about a story relating to Covid 19 vaccination, she noticed that the background of the Provenance icon would sometimes turn red. When she clicked on the icon, the verification criteria information pane showed that Provenance had detected a problem with the image used in the news article she was reading. Clicking on the arrow to open the drop-down explanation pane, she reads that Provenance has detected that the image has been used before in another article. The image in question shows a picture taken at a conference of the World Health Organisation. Looking closely, she sees a credit to the Associated Press (AP). She knows that AP is an international news wire service, and that local and national news agencies republish their articles, including the images. As this is just an image of a press conference, she is confident that its use by multiple news agencies is not an issue.

\section{Evaluation} \label{Evaluation}
Provenance is under development and will shortly be undergoing human evaluation. Currently, five of the seven news analysis functions have been implemented and have been integrated with the platform. These are undergoing technical evaluation while the final two analysis tools are being completed. When the tool is fully completed, a series of technical tests and human evaluation tests will be undertaken to evaluate basic functionality and to ensure that it is providing the right warnings at the appropriate time. Following this, a series of experiments will be undertaken to evaluate its effect on user behaviour. This will include the likelihood of reading and sharing news articles that have cautionary warnings beside them. We will also be analysing unintended effects of the tool. Finally, a series of long term studies are planned to evaluate its effect on users' media literacy.

\section{Conclusions} \label{Conclusions}
Misinformation and disinformation are significant issues that have negatively affected public discourse, politics and social cohesion. The Internet and especially social media are the primary conduits for its growth and spread. Existing user-orientated browser plugins have limited capabilities and only provide users with an historical rating of a website's propensity to publish misinformation and disinformation. They are also not capable of detailed analysis of the content of news webpages or social media feeds. The Provenance browser plugin significantly improves upon existing user orientated solutions by providing intermediary free analysis of webpage and social media content using seven criteria, and where necessary providing cautionary warnings to users. The user can then check the detailed explanatory warning notifications to make their own judgement. This will improve users' media literacy and reduce susceptibility to misinformation and disinformation long term.

\section{Acknowledgements}
The work has been supported by the PROVENANCE project which has received funding from the European Union’s Horizon 2020 research and innovation programme under Grant Agreement No. 825227, and with the financial support of Science Foundation Ireland under Grant Agreement No. 13/RC/2106\_P2 at the ADAPT SFI Research Centre.

\bibliography{main}

\end{document}